\documentclass{article}
\usepackage{graphicx}
\usepackage{amsthm}
\usepackage{latexsym}

\def\be{\begin{equation}}
\def\ee{\end{equation}}

\long\def\symbolfootnote[#1]#2{\begingroup%
\def\thefootnote{\fnsymbol{footnote}}\footnote[#1]{#2}\endgroup}

\newlength{\defbaselineskip}
\newcommand{\setlinespacing}[1]%
           {\setlength{\baselineskip}{#1 \defbaselineskip}}

\newcommand{\singlespacing}{\setlength{\baselineskip}{\defbaselineskip}}

\theoremstyle{plain}
\newtheorem{thm}{Theorem}[section]

\newtheorem{prop}[thm]{Proposition}
\theoremstyle{definition}
\newtheorem{defn}[thm]{Definition}

\theoremstyle{remark}

\begin{document}

\begin{center} {\LARGE\textbf{Quantum observer and Kolmogorov complexity: a model that can be tested}}
\vskip 2em
{\large \bf Alexei Grinbaum} \\
{\it CEA-Saclay/LARSIM, 91191 Gif-sur-Yvette, France
\par Email alexei.grinbaum@cea.fr}
\vskip 1em \today
\end{center}
\vskip 1em
\begin{abstract}\noindent
Different observers do not have to agree on how they identify a  quantum system. We explore a condition based on algorithmic complexity that allows a system to be described as an objective ``element of reality''. We also suggest an experimental test of the hypothesis that any system, even much smaller than a human being, can be a quantum mechanical observer.
\end{abstract}


\section{Introduction}

Quantum mechanical formalism has an orthodox interpretation that relies on the cut between the observer and the system observed~\cite{dirac,vN32}. This ``shifty split''~\cite{Bell3} of the world into two parts cannot be removed: the formalism only applies if the observer and the system are demarcated as two separate entities. Standard quantum mechanics says nothing about the physical composition of the observer, who is an abstract notion having no physical description from within quantum theory. One cannot infer from the formalism if the observer is a human being, a machine, a stone, a Martian, or the whole Universe. As emphasized by Wheeler, this makes it extraordinarily difficult to state clearly where ``the community of observer-participators'' begins and where it ends~\cite{wheelerLaw}.

As a part of his relative-state interpretation, Everett argued that observers are physical systems with memory, i.e., ``parts... whose states are in correspondence with past experience of the observers''~\cite{everett}. We call this a \textit{universal observer} hypothesis: any system with certain information-theoretic properties can serve as quantum mechanical observer, independently of its physical constituency, size, presence or absence of conscious awareness and so forth. In this vein, Rovelli claimed that observers are merely systems whose degrees of freedom are correlated with some property of the observed system: ``Any system can play the role of observed system and the role of observing system. \ldots The fact that observer $O$ has information about system $S$ (has measured $S$) is expressed by the existence of a correlation\ldots''~\cite{RovRQM}. However, the universal observer hypothesis has remained a controversial statement to this day. For example, Peres claims in the way exactly opposite to Rovelli's, that ``the two electrons in the ground state of the helium atom are correlated, but no one would say that each electron `measures' its partner''~\cite{PeresMeasurement}. The purpose of this paper is to clarify an information-theoretic definition of quantum mechanical observer and to propose a physical test of this hypothesis.

In section~\ref{sectSIA} we give a general definition of observer based on the intuitive feeling that a key component of observation is system identification. Then we apply to it the notion of Kolmogorov complexity, which is the main tool of ensuing analysis. In section~\ref{sectQuantClassSystems} this approach is developed to germinate a definition of quantum and classical systems. In section~\ref{sectElementsReality} we consider a family of observers and require that a system be identified by them in the same way, thereby giving an information-theoretic criterion of an ``element of reality''. In section~\ref{sectRelativity} we show that observers can be allowed some disagreement while still maintaining an unambiguous identification of the observed element of reality. Finally, in Section~\ref{ExpSection} we suggest an experimental test of the universal observer hypothesis.

\section{Observer as a system identification algorithm}\label{sectSIA}

What characterizes an observer is that it has information about some physical system. This information fully or partially describes the state of the system. The observer then measures the system, obtains further information and updates his description accordingly. Physical processes listed here: the measurement, updating of the information, ascribing a state, happen in many ways depending on the physical constituency of the observer. The memory of a computer acting as an observer, for instance, is not the same as human memory, and measurement devices vary in their design and functioning. Still one feature unites all observers: that whatever they do, they do it to a \emph{system}. In quantum mechanics, defining an observer goes hand in hand with defining a system under observation. An observer without a system is a meaningless nametag, a system without an observer who measures it is a mathematical abstraction. What remains constant throughout measurement is the identification, by the observer, of the quantum system.

Quantum systems aren't like sweets: they don't melt. Take a general thermodynamic system interacting with other systems. Such a system can dissipate, diffuse, or dissolve, and thus stop being a system. If at first a cube of ice gurgling into tepid water is definitely a thermodynamic system, it makes no sense to speak about it being a system after it has dissolved. Quantum systems aren't like this. Its state may evolve, but the observer knows how to tell the system he observes from the rest or the environment. An electron in a certain spin state remains an electron after measurement, i.e., it remains a system with a particular set of the degrees of freedom. The observer maintains the identity of the system notwithstanding a change in the state of this system that may or may not occur. So, whatever else he might happen to be, the observer is primarily a system identification machine. Different observers having different features (clock hands, eyes, optical memory devices, internal cavities, etc.) all share this central characteristic.

\begin{defn}
An observer is a system identification algorithm (SIA).
\label{defobserver}\end{defn}

Particular observers can be made of flesh or, perhaps, of silicon. `Hardware' and `low-level programming' are different for such observers, yet they all perform the task of system identification. This task can be defined as an algorithm on a universal computer, e.g., the Turing machine: take a tape containing the list of all degrees of freedom, send a Turing machine along this tape so that it puts a mark against the degrees of freedom that belong to the quantum system under consideration. Any concrete SIA may proceed in a very different manner, yet all can be modelled with the help of this abstract construction.

The SIAs whose physical realization may differ share one property that does not depend on the hardware: their algorithmic, or Kolmogorov, complexity. Any SIA can be reconstructed from a binary string of some minimal length (which is a function of this SIA) by a universal machine. As shown by Kolmogorov~\cite{kolmogorov65}, this minimal compression length defines the amount of information in the SIA and does not depend (up to a constant) on the realization of the SIA on this or that hardware. The common-lore view of a multitude of individual observers, one hastily printing, another yawning, a third one moving around his DNA strands, is now superseded by the view of different SIAs, each with its algorithmic complexity defined via a universal machine.

\section{Quantum and classical systems}\label{sectQuantClassSystems}

Each quantum systems has a certain number of the degrees of freedom, which we think about as being independent parameters needed in order to characterize the state of the system. For example, a system with only two states (spin-up and spin-down) has one degree of freedom and can be described by one parameter $\sigma=\pm 1$. If we write down these parameters as a binary string, the Kolmogorov complexity of that string is at least the number of the degrees of freedom of the system. Consequently, for any system $S$, the Kolmogorov complexity of the binary string $s$ representing its parameters
\begin{equation}
K(s) \geq \mathrm{d} _S,
\end{equation}
where $\mathrm{d} _S$ is the number of the degrees of freedom in $S$. In what follows the notation $K(s)$ and $K(S)$ will be used interchangeably.

When we say that an observer $X$ observes a quantum system $S$, it is usually the case that $K(S)\ll K(X)$. In this case the observer will have no trouble keeping track of all the degrees of freedom of the system; in other words, the system will not `dissolve' or `melt' in the course of dynamics. However, it is also possible that $X$ identifies a system with $K(S)>K(X)$. For such an observer, the identity of system $S$ cannot be maintained and some degrees of freedom will fall out from the description that $X$ makes of $S$.

\begin{defn}\label{defquantum}
System $S$ is called quantum with respect to observer $X$ if $K(S)<K(X)$, meaning that $X$ can give a complete description of all its degrees of freedom. Otherwise $S$ is called classical with respect to $X$.
\end{defn}

Suppose that $X$ observes a quantum system $S$ and another observer $Y$ observes both $S$ and $X$. If $K(Y)$ is greater than both $K(X)$ and $K(S)$, observer $Y$ will identify both systems as quantum systems. In this case $Y$ will typically treat the interaction between $X$ and $S$ as an interaction between two quantum systems. If, however, $K(X)$ and $K(Y)$ are close, $K(X)\gg K(S)$ and $K(Y)\gg K(S)$ but $K(X)\simeq K(Y)$, then $Y$ will see $S$ as a quantum system but the other observer, $X$, as a classical system. An interaction with a classical system, which we usually call `observation', is a process of decoherence that occurs when the Kolmogorov complexity of at least one of the systems involved approaches the Kolmogorov complexity of the external observer. In this case $Y$ cannot maintain a complete description of $X$ interacting with $S$ and must discard some of the degrees of freedom. If we assume that all human observers acting in their SIA capacity have approximately the same Kolmogorov complexity, this situation may provide an explanation of the fact that we never see a human observer (or, say, a cat) as a quantum system.

One welcome consequence of Definition~\ref{defobserver} is that Kolmogorov complexity $K(X)$ is not computable. We as human observers do not seem to know the maximum number of the degrees of freedom in a system that we can still keep track of. A photon is certainly a quantum system from our point of view, a simple atom too, a $C_{60}$ perhaps as well, albeit seeing quantum effects with fullerenes is not easy. But we have never seen a quantum system having, say, $10^{23}$ degrees of freedom. So where does the border run? Is it a number like $6$ or $20$ or is it $10^{n},\,n>2$? All we can say is that mathematics shows that human observers cannot compute their own $K(X)$.

\section{Elements of reality}\label{sectElementsReality}

Ever since the EPR paper~\cite{EPR}, the question of what is real in the quantum world has been at the forefront of all conceptual discussions about quantum theory. Einstein, Podolsky and Rosen formulated their question with regard to physical \textit{properties}: e.g., is position or momentum real? This is however not the only problem of reality that appears when many observers enter the game. Imagine a sequence of observers $X_i,\, i=1,\, 2,\ldots$, each identifying systems $S_n,\, n=1,\, 2,\ldots$. System identifications of each $S_n$ do not have to coincide as some observers may have their Kolmogorov complexity $K(X_i)$ below, or close to, $K(S_n)$, and others much bigger than $K(S_n)$. If there is disagreement, is it possible to say that the systems are real, or objects of quantum mechanical investigation, in some sense? We can encode the binary identification string produced by each observer in his SIA capacity as some random variable $\xi _i\in \Omega$, where $\Omega$ is the space of such binary identification strings, possibly of infinite length. Index $i$ is the number of the observer, and the values taken by random variable $\xi_i$ bear index $n$ corresponding to ``$i$-th observer having identified system $S_n$''. Adding more observers, and in the limit $i \rightarrow \infty$ infinitely many observers, provides us with additional identification strings. Putting them together gives a stochastic process $\{ \xi _i \}$, which is an observation process by many observers. If systems $S_n$ are to have a meaning as ``elements of reality'', it is reasonable to require that this stochastic process have entropy rate equal to zero:
\begin{equation}
H (\{ \xi _i\})=0\label{entropyzero}
.\end{equation} We also consider this process stationary and ergodic.

Let us illustrate the significance of condition~(\ref{entropyzero}) on a simplified example. Suppose that $\theta_1, \theta_2,\ldots$ is a sequence of independent identically distributed random variables taking their values among binary strings of length $r$ with probabilities $q_k$, $k\leq 2^r$. These $\theta_k$ can be seen as identifications, by different SIAs, of different physical systems, i.e., a special case of the $\xi_i$-type sequences having fixed length and identical distributions. For instance, we may imagine that $\theta _1$ is a binary encoding of the first observer seeing an electron and $\theta _2$ is a binary string corresponding to the second observer having identified a physical system such as an elephant; and so forth. Entropy becomes simply:
\begin{equation}\label{simplentropy}
H=-\sum _k q_k \log q_k.
\end{equation}
Condition~(\ref{entropyzero}) applied to entropy~(\ref{simplentropy}) means that all observers output one and the same identification string of length $r$, i.e., all SIAs are identical.
This deterministic system identification, of course, obtains only under the assumption that the string length is fixed for all observers and their random variables are identically distributed, both of which are not plausible in the case of actual quantum mechanical observers. So rather than requiring identical strings we impose condition~(\ref{entropyzero}) as a criterion of the system being identified in the same way by all observers, i.e., it becomes a candidate quantum mechanical ``object of investigation''.

\section{Relativity of observation}\label{sectRelativity}

Let us explore the consequences of condition~(\ref{entropyzero}). Define a binary sequence $\alpha ^i _n$ as a concatenation of the system identifications strings of systems $S_n$ by different observers:
\begin{equation}
\alpha ^i _n = \overline{(\xi _1)_n}\,\overline{(\xi _2)_n}\ldots \overline{(\xi _i)_n},
\end{equation}
where index $i$ numbers observers and the upper bar corresponds to ``string concatenation'' (a detailed definition can be found in~\cite{zl}). Of course, this concatenation is only a logical operation and not a physical process. A theorem by Brudno~\cite{brudno78,brudno83} conjectured by Zvonkin and Levin~\cite{zl} affirms that the Kolmogorov complexities of strings $\alpha ^i _n$ converge towards entropy:
\begin{equation}\label{brudno}
    \lim _{n\rightarrow \infty} \lim _{i\rightarrow \infty} \frac{K(\alpha _n ^i)}{i} = H(\{ \xi _i\}).
\end{equation}
For a fixed $i$ and the observer $X_i$ who observes systems $S_n$ that are quantum in the sense of Definition~\ref{defquantum}, variation of $K(\alpha _n ^i)$ in $n$ is bounded by the observer's own complexity in his SIA capacity:
\begin{equation}
    K(\alpha _n ^i) < K(X_i) \qquad \forall n, \qquad i\;\mathrm{fixed}.
\end{equation}
Hence eqs. (\ref{entropyzero}) and (\ref{brudno}) require that
\begin{equation}\label{eqlog}
    \lim _{i\rightarrow\infty} \frac{K(\alpha _n ^i)}{i} =0.
\end{equation}
This entails that the growth of $K(\alpha _n ^i)$ in $i$ cannot be faster than logarithmical.
Therefore the following:
\begin{prop}
An element of reality that may become an object of quantum mechanical investigation can be defined only with respect to a class of not very different observers.
\label{propobs}\end{prop}

To give an intuitive illustration, imagine adding a new observer $X_{i+1}$ to a group of observers $X_1,\ldots,X_i$ who identify systems $S_n$. This adds a new identification string that we glue at the end of concatenated string $\alpha _n ^{i}$ consisting of all $X_i$'s identifications of $S_n$, thus obtaining a new string $\alpha _n ^{i+1}$. The Kolmogorov complexity of $\alpha _n ^{i+1}$ does not have to be the same as the Kolmogorov complexity of $\alpha _n ^{i}$; it can grow, but not too fast, i.e., not faster than the logarithm. Adding a new observation may effectively add some new non-compressible bits, but not too many such bits. If this is so, then $H=0$ still obtains. Although observers $X_1,\ldots,X_i,X_{i+1}$ produce slightly different identification strings, they agree, simply speaking, that an atom is an atom and not something that looks more like an elephant.

The above reasoning applies only to quantum systems $S_n$ in the sense of Definition~\ref{defquantum}. This is because in the case of classical systems different observers may each operate their own coarse-graining, keeping only some degrees of freedom. System identification strings may then differ dramatically and one cannot expect $K(\alpha _n ^i)$ to grow moderately.

\section{Experimental test}\label{ExpSection}

\begin{figure}
\centering
\includegraphics[width=9cm]{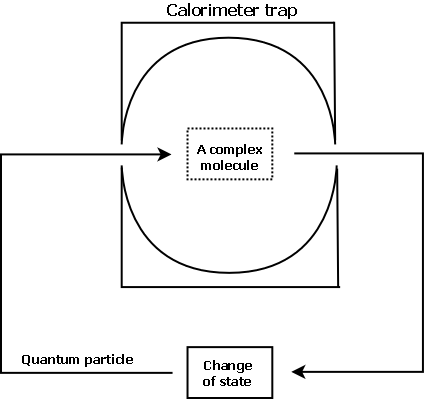}
\caption{Experiment leading to heat production when physical entropy $\mathcal{S}=H+K$ changes.}
\label{Test1}\end{figure}

A previously suggested experimental connection between thermodynamics and theories based on Kolmogorov complexity is based on observing the consequences of a change in the system's state~\cite{Zurek1,Zurek3,Zurek2,Erez}. Zurek introduced the notion of physical entropy $\mathcal{S}=H+K$, where $H$ is the thermodynamic entropy and $K$ the Kolmogorov entropy. If the observer with a finite memory has to record the changing states of the quantum system, then there will be a change in $\mathcal{S}$, like the one depicted on Figure~\ref{Test1}, and it will lead to heat production that can be observed experimentally.

\begin{figure}
\centering
\includegraphics[width=9cm]{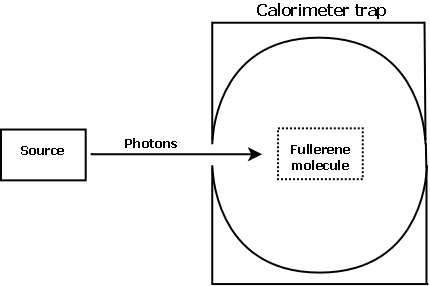}
\caption{Experiment leading to heat production when observer's memory becomes saturated.}
\label{Test2}\end{figure}

What we propose here, based on a suggestion by Anton Zeilinger, is a simpler setting that can still serve as a test of the universal observer hypothesis. It does not rely on the measurement of particular states, but on the fact of measurement as such. If a measurement occurs, then the observer has identified the quantum system, and this fact in itself, if repeated, will eventually lead to heat production.

An individual fullerene molecule is placed in a highly sensitive calorimeter and bombarded with photons, which play the role of quantum systems with low $K(S)$ (Figure~\ref{Test2}). According to the universal observer hypothesis, the fullerene is a quantum mechanical observer with $K(X)>K(S)$. Thus the absorption of the photon by the fullerene can be described as measurement: the fullerene identifies a quantum system, i.e. the photon, and observes it, obtaining new information. Physically, this process amounts to establishing a correlation between the photon variables, i.e., its energy, and the degrees of freedom of the fullerene. The external observer knows that such a process has occurred but remains unaware of its exact content, so that he is aware that there has been measurement, but doesn't know a precise state of the photon as measured by the fullerene, nor a precise state of the fullerene after measurement.

Informationally speaking, the same process can be described at storing information in the fullerene's memory. If measurement is repeated on several photons, more such information is stored, so that at some point total Kolmogorov complexity will approach $K(X)$. When it reaches $K(X)$, the fullerene will stop identifying incoming photons as quantum systems. Any further physical process will lead to heat production due to the erasure of memory, as prescribed by the Landauer principle. Physically, this process will correspond to a change of state of the carbon atoms that make up the fullerene molecule: the calorimeter will register a sudden increase in heat when $C_{60}$ cannot store more information, thereby ending its observer function.

Actual experiments with fullerenes show that this scenario is realistic. A fullerene molecule ``contains so many degrees of freedom that conversion of electronic excitation to vibrational excitation is extremely rapid''. Thus, the fullerene is a good candidate for a quantum mechanical observer, for ``the molecule can store large amounts of excitation for extended periods of time before degradation of the molecule (ionization or fragmentation) is observed''~\cite{Fullerenes2}. The experiments in which fullerenes are bombarded with photons demonstrate that ``the energy of the electronic excitation as a result of absorption of a laser photon by a molecule is rapidly converted into the energy of molecular vibrations, which becomes distributed in a statistical manner between a large number of the degrees of freedom of the molecule\ldots The fullerene may absorb up to 10 photons at $\lambda=308$ nm wavelength before the dissociation of the molecule into smaller carbon compounds''~\cite{Fullerenes1}. We read these results as a suggestion that there should be one order of magnitude difference between $K(S)$ and $K(X)$ and that this allows the fullerene to act as quantum mechanical observer for up to 10 photons at 308 nm wavelength. What needs to be tested experimentally in this setting is heat production: we conjecture that if the same process occurs inside a calorimeter, the latter will register a sudden increase in heat after the fullerene will have observed 10 photons. What we predict here isn't new physics, but an explanation on a new level: that of information, of a physical process: heat production, whose role during the dissociation of fullerenes has been largely overlooked. We suggest that heat production deserves special attention as a signature of the fullerene's role as quantum mechanical observer.

As a side remark, imagine that the photon's polarization state in some basis were fully mixed: $$\frac{1}{2} (|0\rangle + |1\rangle).$$ While only the energy of the photon matters during absorption, the external observer records von Neumann entropy $H=\log 2$ corresponding to this mixture (the initial state of the fullerene is assumed fully known). After absorption, it is mandatory that this entropy be converted into Shannon entropy of the new fullerene state, corresponding nicely to the uncertainty of the external observer in describing the ``statistical manner'' of the distribution over a large number of the degrees of freedom. From the internal point of view, we may assume perfect `self-knowledge' of the observer, which puts his Shannon entropy equal to zero. However, his Kolmogorov entropy will increase as a result of recording the measurement information~\cite{Zurek3}. Heat produced during the erasure of measurement information is at least equal to Kolmogorov complexity of the string that was stored in observer's memory; but according to quantum mechanics, this heat will not reveal to the external observer any information about the precise photon state observed by the fullerene.

\section{Conclusion}

The Copenhagen view of quantum mechanics traditionally described quantum systems and observers, epistemologically, as belonging to different categories. On the contrary, the view based on the relativity of observation, as proposed by Everett and later Rovelli, puts all systems on equal grounds and ascribes them only relative states. These two views are not as contradictory as they may seem. Relativity of observation has been understood by some proponents of the Copenhagen school~\cite{hermann,fock2,fock3}. Information-theoretic treatment of the observer gives a chance to completely overcome the tension. On the one hand, the observer is a SIA and is characterized by its Kolmogorov complexity. On the other hand, quantum mechanics can be reconstructed from information-theoretic axioms and thus seen as a theory of information~\cite{grinbjps}. This puts all systems on equal grounds, in the spirit of Rovelli, while emphasizing the idea of relativity of observation, in the spirit of Fock.

Additionally, information-theoretic treatment of the observer provides a somewhat surprising result developing EPR's notion of ``element of reality''. One can make sense of a system existing independently of observation, with respect to a class of observers whose Kolmogorov complexities may differ, even if slightly. Equation~(\ref{eqlog}) provides a mathematical criterion for this.

We have analyzed the observer as a system identification algorithm in the context of quantum mechanics. It remains an open question to apply this analysis to quantum field theory, where the task of system identification may look significantly different from the finite-dimensional situation. It also remains an open problem to realize experimentally the setup proposed in Section~\ref{ExpSection}, which may lead to experimental confirmation of the universal observer hypothesis. Putting together this experimental test, which may show that a fullerene can act as a observer for up to 10 photons, and the remark on non-computability of $K(X)$ at the end of Section~\ref{sectQuantClassSystems} begs yet another question: is it possible to say that, although $K(X)$ isn't computable in the mathematical sense, physical experiment effectively computes it?

\section*{Acknowledgements}

I am grateful to Vasily Ogryzko for stimulating discussions and to \v Caslav Brukner, Markus Aspelmeyer, Ognyan Oreshkov and Anton Zeilinger for their remarks and hospitality at the Institute for Quantum Optics and Quantum Information of the Austrian Academy of Sciences.

\singlespacing\footnotesize


\end{document}